\documentclass[aps,prd,twocolumn,superscriptaddress,groupedaddress]{revtex4}

\usepackage{dcolumn}   % needed for some tables
\usepackage{bm}        % for math
\usepackage{multirow}

\usepackage{amsmath}
\usepackage{amsfonts}
\usepackage{amssymb}
\usepackage{makeidx}
\usepackage{graphicx}
\usepackage{anysize}

\newcommand{\bs}{\boldsymbol}

\newcommand{\ba}{\begin{eqnarray}}
\newcommand{\ea}{\end{eqnarray}}
\newcommand{\be}{\begin{equation}}
\newcommand{\ee}{\end{equation}}

\usepackage[dvips]{color}

\begin{document}

\title{Parity violation in quasielastic electron-nucleus scattering within the relativistic impulse approximation}

\author{R.~Gonz\'alez-Jim\'enez}
\affiliation{Departamento de F\'isica At\'omica, Molecular y Nuclear, Universidad de Sevilla, 41080 Sevilla, Spain}
\affiliation{Department of Physics and Astronomy, Ghent University, Proeftuinstraat 86, B-9000 Gent, Belgium}

\author{J.A.~Caballero}
\affiliation{Departamento de F\'isica At\'omica, Molecular y Nuclear, Universidad de Sevilla, 41080 Sevilla, Spain}

\author{T.W.~Donnelly}
\affiliation{Center for Theoretical Physics, Laboratory for Nuclear Science and Department of Physics,
Massachusetts Institute of Technology, Cambridge, Massachusetts 02139, USA}

\date{\today}

\begin{abstract}
We study parity violation in quasielastic (QE) electron-nucleus scattering using the relativistic impulse approximation.
Different fully relativistic approaches have been considered to estimate the effects associated with the final-state interactions. 
We have computed the parity-violating quasielastic (PVQE) asymmetry and have analyzed its sensitivity to the different ingredients that
enter in the description of the reaction mechanism: final-state interactions, nucleon off-shellness effects, current gauge ambiguities.
Particular attention has been paid to the description of the weak neutral current form factors.
The PVQE asymmetry is proven to be an excellent observable when the goal is to get precise information on the axial-vector sector 
of the weak neutral current. Specifically, from measurements of the asymmetry at backward scattering angles good knowledge of the 
radiative corrections entering in the isovector axial-vector sector can be gained.
Finally, scaling properties shown by the interference $\gamma-Z$ nuclear responses are also analyzed. 
\end{abstract}

% \pacs{}
\maketitle

\tableofcontents

\section{Introduction}

% PV in electron scattering
%%%%%%%%%%%%%%%%%%%%%%%%%%%%%%%%%

As is well known, the purely electromagnetic (EM) interaction clearly dominates electron scattering reactions, being 
Parity-Conserving (PC) processes. However, the electron also interacts through the weak neutral current (WNC) interaction 
that does not conserve parity, {\it i.e.,} via Parity-Violating (PV) processes.
Although the weak interaction is several orders of magnitude smaller than the EM one, the role played by the former 
in the scattering process can shed some light on specific ingredients in the reaction mechanism that are not accessible from 
studies considering only the EM interaction.
The main objectives pursued through the analysis of PV electron scattering reactions are:
(i) to serve as a test of the Standard Model,
(ii) to serve as a tool to determine the electroweak form factors of the nucleon, and
(iii) to use the weak interaction as a probe to study nuclear structure.
In this work we focus on the second goal, that is, getting information on the nucleonic structure.

The measurement of PV effects in electron-nucleon/nucleus scattering requires one to build observables that show
a very high sensitivity to the electroweak interaction and which are insensitive to contributions arising from the dominant EM force. 
The helicity asymmetry or PV asymmetry is defined as the ratio between the difference and sum of 
cross sections with opposite helicity of the incident electron,
\ba
{\cal A}^{PV} = \frac{\sigma^+ - \sigma^-}{\sigma^+ + \sigma^-}=\frac{\sigma^{PV}}{\sigma^{PC}}\,,\label{Asym}
\ea
where the superscript $+/-$ denotes positive/negative helicity.
The numerator in Eq.~(\ref{Asym}) is only different from zero if PV effects are considered. 
Hence we denote the resulting cross section as $\sigma^{PV}$. 
On the contrary, the denominator in Eq.~(\ref{Asym}) is dominated by the PC EM interaction, 
so it is denoted as $\sigma^{PC}$.
It is important to point out that previous comments refer to inclusive processes in which the target nucleus is not polarized. 
In any other situation, nuclear responses linked only to the EM interaction can also contribute to the numerator in the 
PV asymmetry~\cite{Raskin88,Donnelly86,exclusive}.

\begin{figure}[htbp]
    \centering
        \includegraphics[width=.35\textwidth,angle=0]{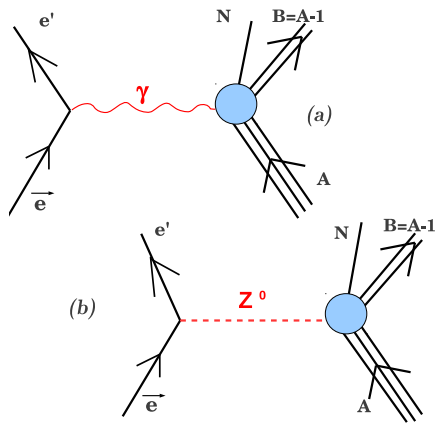}
    \caption{(Color online) Feynman diagrams representing PVQE electron-nucleon scattering in Born approximation: 
      (a) EM interactioon -- one photon ($\gamma$) exchanged, (b) WNC interaction -- one $Z^0$ exchanged.}
    \label{fig:diagrams}
\end{figure}

In this work we restrict our attention to the study of inclusive PV electron-nucleus 
scattering processes, $A(\vec{e},e')B$. 
We consider the QE regime that corresponds to the electron being scattered from a single nucleon that is subsequently ejected 
from the target nucleus. Within the Born approximation, in which the interaction is described by the exchange of a single virtual boson, 
the Feynman diagrams that represent the scattering process are those represented in Fig.~\ref{fig:diagrams}: 
(a) one-photon exchange (EM interaction) and (b) one-$Z^0$ exchange (WNC interaction).
The first-order contribution to $\sigma^{PV}$ in Eq.(~\ref{Asym}) arises from the interference between diagrams 
 (a) and (b) in Fig.~\ref{fig:diagrams}. 
The helicity asymmetry, denoted also as the PVQE asymmetry, ${\cal A}^{PV}_{QE}$, provides a very useful tool to study the WNC interaction. 
For some specific kinematical conditions, that are analyzed in Sect.~\ref{results}, the PVQE asymmetry can help in determining the isovector contribution in the axial-vector form factor of the nucleon. 
This aspect of the problem was already suggested in some previous work~\cite{Donnelly92,Musolf92,Horowitz93a,Musolf94}, and it
can complement the whole set of information provided by the analysis of parity violation in elastic electron-proton scattering.
This latter process was studied in detail in~\cite{Gonzalez-Jimenez13a,Gonzalez-Jimenez14a,Moreno14} with emphasis on the vector strange form factors of the nucleon and the effects linked to the radiative corrections. 
These ingredients also play a role in the case of QE electron scattering. Furthermore, the complex structure of the nuclear target
introduces additional difficulties that should be carefully addressed. 
Some questions of importance in describing PVQE electron scattering are:
\begin{itemize}
 \item Do the current nuclear models reproduce PC processes with enough precision to be used for PV processes? 
 %%%
 \item Which channels beyond the impulse approximation (IA) contribute to the PVQE asymmetry?
 %%%
 \item How important are the modifications of the nucleon form factors due to the nuclear 
   medium and to their off-shell character?
 %%%
 \item What role is played in the PV asymmetry by the Coulomb distortion of the electron wave functions?
 %%%
\end{itemize}
With regards to the first question, the relativistic model considered in this work (relativistic mean field model, RMF) 
has been widely and successfully tested in several previous studies~\cite{Caballero06,Meucci09}. 
Within the RMF final-state interactions (FSI) between the ejected nucleon and the residual nucleus are incorporated using 
the same mean field employed in describing the wave functions of the bound nucleons (we refer to it as the RMF-FSI model).
Particular mention should be drawn to the phenomenon of scaling and the excellent description of the experimental 
data~\cite{Caballero06} provided by the model. 
Contrary to most non-relativistic models, the RMF-FSI is capable of describing not only the magnitude of the experimental scaling function, but also its asymmetric shape 
and the fact that is has a long tail extending to large values of the transferred energy.
On the other hand, the PVQE asymmetry, being built as a ratio of cross sections, is expected to be only mildly modified 
by the uncertainties linked to the nuclear models.

In some previous work~\cite{Amaro02,Alberico93,Barbaro94} it has been 
shown that effects connected to two-body currents are small in the transverse responses attached to the interference between EM and WNC currents (in what follows denoted simply as PV responses); 
however, the situation is clearly different in the longitudinal channel. Here effects beyond the IA can be very significant.
This result can make it difficult to get information on the nucleonic structure from PVQE asymmetry data taken at 
forward scattering angles (where the longitudinal response is not negligible).
On the contrary, at backward angles the longitudinal contribution is tiny (see Sect.~\ref{results}), and therefore, 
effects beyond the IA lead to very mild changes in the PVQE asymmetry.

The potential modification of the form factors of the nucleon due to the nuclear medium has been studied in some 
previous work considering various theoretical approaches: 
the quark-meson coupling model (QMC)~\cite{Lu98,Lu99} and the light-front constituent quark model~\cite{Frank96}.
Both models provide expressions for the EM form factors that depend on the nuclear density.  
In ~\cite{Martinez04,Strauch03,Cristina2004} results for polarization observables corresponding to the exclusive 
process $A(\vec{e},e'\vec{p})B$ were computed using the form factor prescription given by QMC. 
These results do not differ significantly from those computed using the {\it free} prescription of the form factors. 
Thus in this work all results are computed using the free prescription for the nucleon form factors.
On the other hand, concerning the strange form factors, Horowitz and Piekarewicz pointed out in ~\cite{Horowitz93b} that the strangeness content in the nucleon is expected to increase in a significant way with the nuclear density. 
Nevertheless, these results have not been confirmed nor has realistic modeling of such effects yet been developed. 
Consequently, in this work we assume that the strange matrix elements in the nucleon do not depend on the nuclear density.

The approximations used to deal with the off-shell vertex are discussed in Sect.~\ref{formalism}.
This subject has been treated in detail in~\cite{Caballero93,Martinez02a,Naus87,Tie87} in the case of PC electron scattering reactions. 
Results shown in this work complement the more elaborated study presented in our companion paper~\cite{exclusive} where the focus is placed on
exclusive $(\vec{e},e'N)$ reactions. 
Although the latter are not adequate to analyze PV effects due to
the presence of the so-called fifth EM response function, we have considered its analysis to be of interest in order
to get some insight concerning the off-shell and gauge ambiguities in the PV observables. 

Finally, we briefly address the question of the Coulomb distortion of electrons.
This aspect of the problem has been analyzed in previous work using non-relativistic 
approaches~\cite{Giusti87,Spaltro93} as well as a fully relativistic description~\cite{Udias1993,Udias93}.
The incorporation of Coulomb effects introduces important complications in the treatment of the scattering process. 
Not only does the required computational time explode, but also the clear separation between the leptonic and hadronic tensors 
with the subsequent appearance of the response functions does not work any more.
On the other hand, the heavier the target is (and/or the lower the energy of the incident electrons is), the larger 
the effects introduced by the Coulomb distortion are. 
In this work we restrict our study to relatively light nuclei $^{12}$C and $^{16}$O (and $^{40}$Ca in a few cases), and high energies. 
Therefore, all results in this work have been computed within the plane-wave Born approximation (PWBA), {\it i.e.,} a single virtual 
exchanged boson is responsible for the electron-nucleon interaction and the wave functions of incident and scattered electrons 
are described as Dirac plane waves.
%%%%
%%%%
%%%%

In what follows we summarize how this work is organized. In Sect.~II we present the basics of the general formalism involved in the
description of PVQE electron-nucleus scattering. Here we introduce the approaches considered in this work as well as a brief discussion of
the WNC nucleon form factors. In Sect.~III we present and discuss our results for the PV
nuclear responses and the PVQE electron helicity. Here we consider different kinematical situations and examine in detail the effects
associated with the description of FSI, relativistic dynamics, off-shellness and weak nucleon form factors. In Sect.~IV we apply 
scaling arguments to the PV nuclear responses by constructing PV scaling functions to be compared with the EM ones. Finally, in 
Sect.~V we summarize our basic findings and discuss our conclusions.

\section{Formalism}\label{formalism}

In this section we present the basic formalism involved in the description of PVQE electron-nucleus scattering.
Here a longitudinally polarized incident electron (characterized by the four-momentum $K_i^\mu=(\varepsilon_i,{\bf k}_i)$) interacts
with the target nucleus that is assumed to be at rest in the laboratory frame ($P_A^\mu=(M_A,{\bf 0})$).
The interaction is described assuming the Born approximation, {\it i.e.,} only one virtual boson 
(photon for the EM interaction and Z for the WNC one) with four-momentum $Q^\mu=(\omega,{\bf q})$
is considered to be exchanged in the process. The scattered electron carries a four-momentum $K_f^\mu=(\varepsilon_f,{\bf k}_f)$
and the residual nuclear system is characterized by $P_B^\mu=(E_B,{\bf p}_B)$.

This work deals with the description of the QE regime and we make use of the impulse approximation (IA), that is, the boson exchanged is attached directly to a 
single nucleon which is then ejected from the nucleus. 
Within this framework the inclusive cross section corresponding to $(\vec{e},e')$
processes is simply given as an incoherent sum of single-nucleon scattering processes. 
Although this is an important simplification in the description of the electron-nucleus scattering mechanism reaction, it has shown its 
validity when applied to the QE kinematical domain. Hence the inclusive differential cross section, 
$d\sigma/(d\Omega_f d\varepsilon_f)$, is built from the exclusive one (see ~\cite{exclusive} for details) by integrating over
the variables of the scattered nucleon ($d\Omega_N=d(cos\theta_N)d\phi$) and summing over the $A$ nucleons in the target nucleus.
The general expression for the $(\vec{e},e')$ differential cross section can be written in terms of nuclear response functions as:
\ba
\dfrac{d\sigma}{d\varepsilon_f d\Omega_f} 
&=& \sigma_{Mott} \biggl\lbrace  
v_L R^{L} + v_T R^{T}\nonumber\\
%%%%%%%%%%%%%%%%%%%%%
&-& \frac{{\cal A}_0}{2}\left[ (a_V-ha_A) 
\left( v_L\widetilde{R}^{L}+
v_T\widetilde{R}^{T}\right)\right.\nonumber\\
%%%%%%%%%%%%%%%%%%%%%
&+&\left. (ha_V-a_A) v_{T'}\widetilde{R}^{T'}\right]\biggr\rbrace \label{SEDRFG} \,,
\ea
where $\sigma_{Mott}$ is the Mott cross section and $v_\alpha$ the usual lepton kinematical factors 
(see~\cite{Day90} for the explicit expressions). The functions
$R^{L,T}$ ($\widetilde{R}^{L,T}$) are the longitudinal ($L$) and transverse ($T$) EM (PV) nuclear responses 
while $\widetilde{R}^{T'}$ is the PV transverse-axial nuclear response. 
%The labels $L$ and $T$ refer to the direction of the momentum transferred ${\bf q}$.
We have also introduced the function ${\cal A}_0=G_F|Q^2|/(2\sqrt{2}\pi\alpha)$
that sets the scale of the PV cross section.
$G_F$ is the Fermi coupling and $\alpha$ the fine structure constant.
Finally, $a_A=-1$ and $a_V=-1+4\sin^2\theta_W$ represent the vector and axial-vector WNC electron couplings, 
$\theta_W$ being the weak mixing angle. 

\begin{figure}[htbp]
    \centering
        \includegraphics[width=.4\textwidth,angle=0]{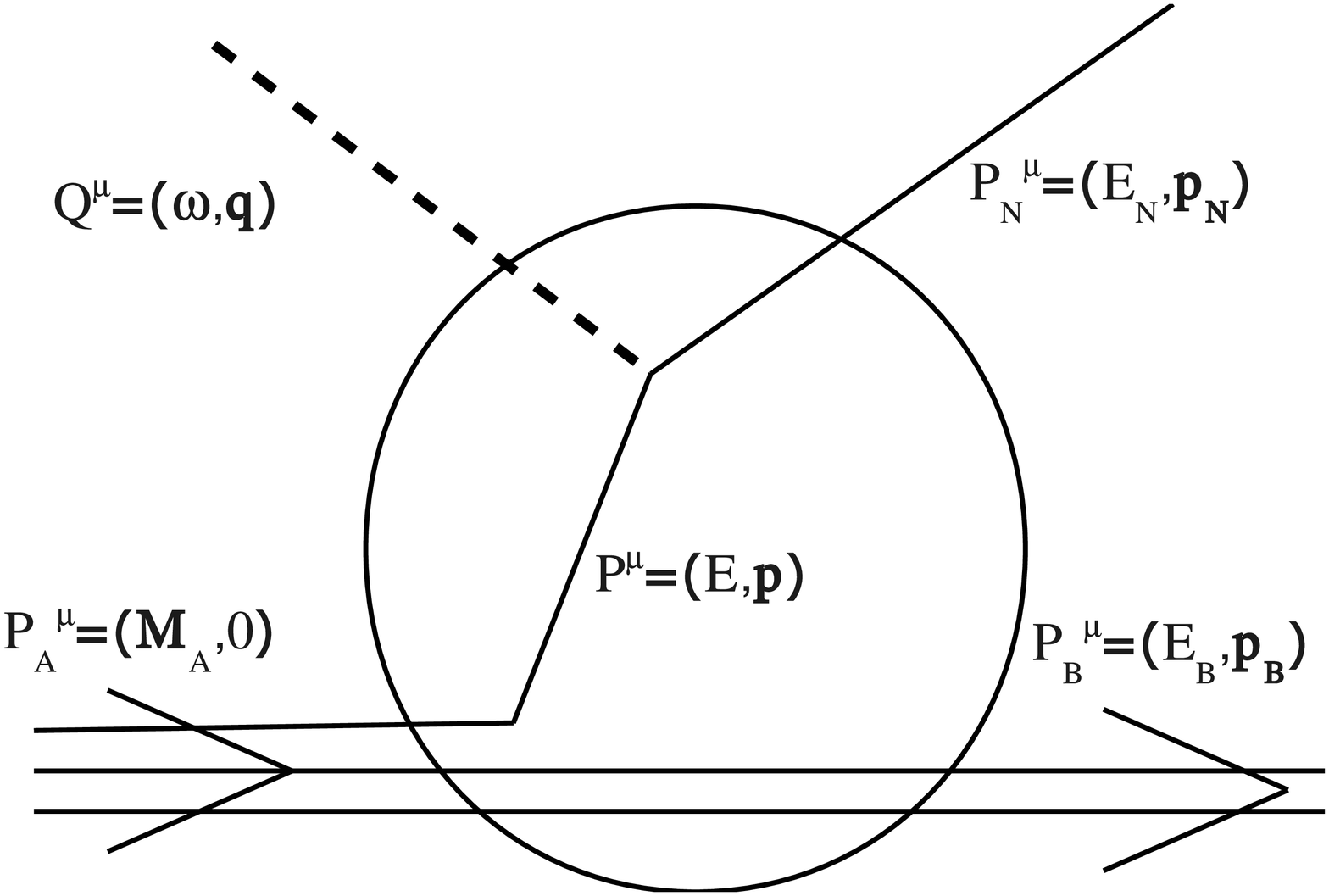}
    \caption{Hadronic vertex in RPWIA.}
    \label{fig:IA_PW}
\end{figure}

Within the IA the nuclear current operator, $\hat{J}^\mu({\bf q})$, is taken as a one-body operator. 
In momentum space the current matrix element can be simply written as
\be
J^\mu\equiv\int d{\bf p}\ \overline{\Phi}_F({\bf p}+{\bs q})
    \hat{J}^\mu\Phi_B({\bf p})\, ,\label{JFIp} 
\ee
where $\Phi_B$ ($\Phi_F$) is the bound (scattered) nucleon wave function and $\hat{J}_N^\mu$ is the one-body current operator. 
Here we make use of the RMF model, presented in detail in ~\cite{Walecka74,Serot86,Serot92,Ring96}, to describe the bound nucleon wave functions. 
On the contrary, different aproaches are considered for the nucleon scattered wave function. First we assume the
relativistic plane-wave impulse approximation (RPWIA). Here the outgoing nucleon, $\Phi_F$, is described by a 
relativistic (four-component) plane wave. This approach is schematically represented in Fig.~\ref{fig:IA_PW}. 
The virtual boson transfers its energy, $\omega$, and momentum, ${\bf q}$, to one nucleon (characterized by $P^\mu=(E,{\bf p})$)
that is subsequently ejected from the nucleus with a four-momentum $P_N^\mu=(E_N,{\bf p}_N)$.

\begin{figure}[htbp]
    \centering
        \includegraphics[width=.4\textwidth,angle=0]{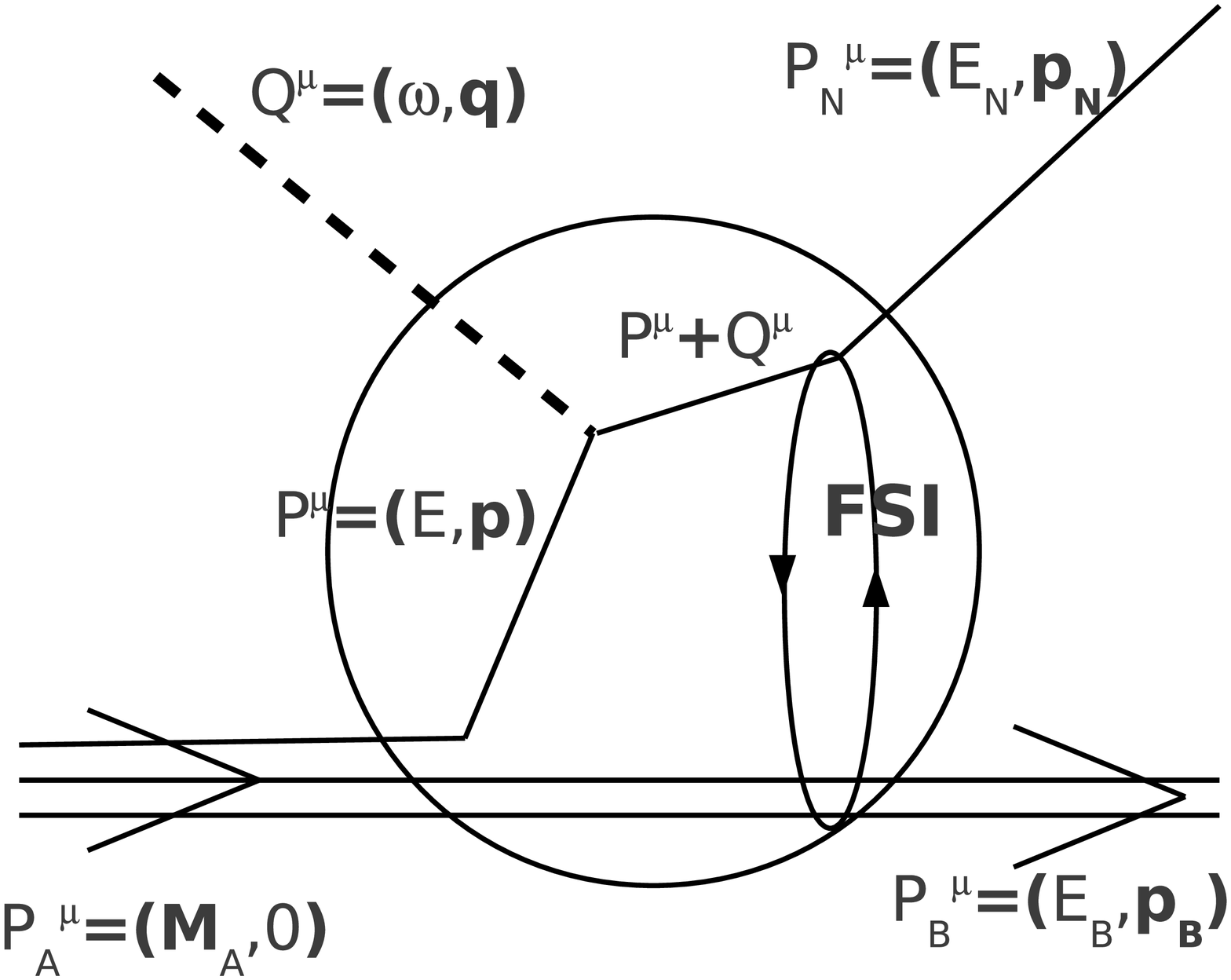}
    \caption{Hadronic vertex when FSI are considered (RMF-FSI model).}
    \label{fig:IA_FSI}
  \end{figure}

Final-state interactions (FSI) are an essential ingredient in describing electron-nucleus scattering. Hence, we incorporate FSI in our 
model by describing the wave function of the outgoing nucleon, $\Phi_F$, as a scattered solution of the Dirac equation in presence 
of the same scalar ($S$) and vector ($V$) potentials employed in the description of the bound wave function. Contrary to the 
complex phenomenological potentials used for $(\vec{e},e'N)$ processes (see \cite{exclusive}), the real potential considered in our
present case (inclusive electron scattering) preserves the flux. Moreover, the use of the same relativistic potential for both the
bound and scattered wave functions is consistent with having the continuity equation fulfilled.
Figure~\ref{fig:IA_FSI} shows schematically the situation when FSI are considered. Note the difference between the four-momentum
acquired by the nucleon attached to the boson ($P^\mu+Q^\mu$) and the asymptotic (final-state) value: $P_N^\mu=(E_N,{\bf p}_N)$.

With regards to the current operator we follow the usual procedure, originally proposed in ~\cite{Forest83} and widely 
used in many later works (see~\cite{Caballero98a,Caballero98b} and references therein), that consists of taking the current operator corresponding to free (on-shell) nucleons. 
Explicit expressions have been shown in numerous previous works~\cite{Boffi93,Caballero93,Caballero98a,Martinez02b} and here we make use of the notation introduced in our accompanying work~\cite{exclusive}. 
Following with the general discussion presented in \cite{exclusive}, the use
of the two usual prescriptions, CC1 and CC2, that are totally equivalent for on-shell nucleons, leads to different results in the
case of bound/scattered, {\it i.e.,} off-shell, nucleons. Moreover, current conservation (for the EM and vector weak currents)
is not fulfilled, hence results may differ significantly when different gauges are selected. We consider the three usual
options: 
i) Landau (no current conservation imposed, denoted as NCC1/NCC2), 
ii) Coulomb (vector current conservation is restored by using $J^3=(\omega/q) J^0$, denoted as CC1(0) and CC2(0)), and 
iii) Weyl (current conservation imposed by $J^0=(q/\omega) J^3$, denoted as CC1(3) and CC2(3)). 
In the next section a detailed study of the various PV inclusive
responses and the PV asymmetry is presented with emphasis on their sensitivity with FSI, off-shellness and 
the choice of current operator.

\subsection{WNC nucleon form factors}\label{WNCff}

The PVep asymmetry (defined for elastic electron-proton scattering) provides an excellent tool to get information on the electroweak structure of the proton. 
This subject was investigated in detail in ~\cite{Gonzalez-Jimenez13a}. 
Here our interest is focused on the analysis of
PVQE electron-nucleus scattering. This process may provide information on the nucleon structure that can complement what is obtained 
from the PVep asymmetry, even being aware of the uncertainties linked to nuclear effects. 
In this section we summarize the basic points considered in this work concerning the description of the EM and WNC form factors.
As already mentioned in the introduction, there exist different approaches that provide precise descriptions of the purely
EM nucleon form factors in the QE kinematical region of interest for this work~\cite{Gonzalez-Jimenez13a}. Here we have
considered the GKex model~\cite{Lomon01,Lomon02,Crawford10} to describe the electric and magnetic nucleon form factors: $G_{E,M}^{p,n}$.

With respect to the weak sector, assuming charge symmetry, the WNC nucleon form factors can be expressed as~\cite{Musolf94}:
\ba
 \widetilde{G}_{E,M}^{p,n}
  &=& (1-4\sin^2\theta_W)(1+R_V^p)G_{E,M}^{p,n}\nonumber\\ 
  &-& (1+R_V^n)G_{E,M}^{n,p} - (1+R_V^{(0)}) G_{E,M}^{(s)} \, , \nonumber \\
&&
\label{WGEMpn}
\ea
where $G_{E,M}^{(s)}$ are the electric, $E$, and magnetic, $M$, strange form factors. 
We assume the $Q^2$-dependence of the strange form factors to be described as follows:  
$G_E^{(s)}(Q^2)=\rho_s \tau G_D^V(Q^2)$ 
and $G_M^{(s)}(Q^2)=\mu_s G_D^V(Q^2)$, where 
$G_D^V=(1+|Q^2|/M_V^2)^{-2}$ is the dipole form factor with $M_V=0.84$ GeV.

The parameters $\rho_s$ and $\mu_s$ determine the size of the strange quark contributions to the electric and magnetic 
vector current in the nucleon, respectively. In this work we make use of the results given in ~\cite{Gonzalez-Jimenez13a}: 
$\rho_s=0.59\pm0.62$ and $\mu_s=-0.02\pm0.21$. 
Notice that the previous uncertainties are much larger 
than the ones shown in~\cite{Gonzalez-Jimenez13a} 
but consistent with those shown in~\cite{Gonzalez-Jimenez14a}.
This is due to the particular procedure considered in their evaluation. 
% Once one of the parameters is fixed to the central
% value of the $[\mu_s-\rho_s]$ 1$\sigma$ confidence ellipse shown in Fig.~14 (panel (i)) of ~\cite{Gonzalez-Jimenez13a}, the other
% parameter is varied along the $x$ and/or $y$ axes up to the horizontal ($\rho_s$) or vertical ($\mu_s$) line to
% intersect with the extremes of 
% the confidence ellipse. 
% The values of the parameters obtained in this way are consistent with the ones presented 
% in~\cite{Gonzalez-Jimenez14a}.

The WNC axial-vector form factor can be written as~\cite{Musolf94}:
\ba
 G_A^{eN} &=& \left[-2(1+R_A^{T=1})G_A^{(T=1)}\tau_3
 + \sqrt{3}R_A^{T=0}G_A^{(8)} \right. \nonumber\\ 
 &+&
 \left. (1+R_A^{(0)})G_A^{(s)}\right]G^A_D(Q^2)\label{WGApn} \, ,
\ea
where the label $N$ denotes proton or neutron and the isospin index $\tau_3=1$ ($-1$) for proton (neutron) has been introduced.
The term $G_A^{(T=1)}\equiv g_A = 1.2695$ represents the isovector contribution to the axial-vector form factor while 
$G_A^{(8)}\equiv(3F-D)/(2\sqrt{3})=0.58\pm0.12$ and $G_A^{(s)}\equiv\Delta s = -0.07\pm0.06$ are the octet and strange 
isoscalar contributions. We use the standard dipole shape for the functional dependence of the axial-vector form factor:
$G^A_D(Q^2)=(1+|Q^2|/M_A^2)^{-2}$ with $M_A=1.03$ GeV.

In Eqs.~(\ref{WGEMpn},~\ref{WGApn}) the terms $R$ represent the radiative corrections.
In this work we consider the values presented in ~\cite{Liu07}.
It is important to point out that the main sources of uncertainties in the axial-vector form factor, once one assumes a 
functional $Q^2$-dependence, comes from the radiative corrections; in particular, the corresponding ones 
that enter in the isovector ($T=1$) sector of the axial-vector form factor that constitutes the main contribution to $G_A^{ep}$. 
Following ~\cite{Liu07}, we consider $G_A^{ep}(0)=1.04\pm0.44$ where the large uncertainty comes from the error in 
$R_A^{T=1}=-0.258\pm0.34$. In the next section we analyze in detail the effects of these uncertainties on the PVQE asymmetry.

\section{Results}\label{results}

In this section we perform a detailed analysis of the PVQE asymmetry with the goal of getting additional 
information on the electroweak structure of the nucleon. To that end, it is essential to evaluate the effects linked to
non-nucleonic ingredients, in particular, final-state interactions, off-shell nucleon uncertainties 
and effects from relativistic dynamics. We investigate how these ingredients affect the PVQE asymmetry and compare them with
the ones associated with the particular description of the EM and WNC nucleon form factors.

The PVQE asymmetry corresponding to $(\vec{e},e')$ processes 
can be written in terms of the nuclear response functions as follows:
 \begin{eqnarray}
 {\cal A}^{PV}_{QE} \approx \frac{{\cal A}_0}{2{\cal G}^2} 
 \left[a_A\bigl(v_L\widetilde{R}^L+v_T\widetilde{R}^T\bigr)
 -a_Vv_{T'}\widetilde{R}^{T'}\right]\, , \nonumber \\
& & \label{APVQE}
\end{eqnarray}
where we have defined ${\cal G}^2 \equiv v_LR^L+v_TR^T$. This means that we neglect 
the contribution from the PV responses when summing up the cross sections for both electron helicities, {\it i.e.,}
$\sigma^+ + \sigma^-$. This approach works perfectly well because of the extremely large difference between the magnitudes of
the purely EM and PV response functions, the latter being 4-5 orders of magnitude smaller. 

\begin{figure}[htbp]
    \centering
        \includegraphics[width=.34\textwidth,angle=270]{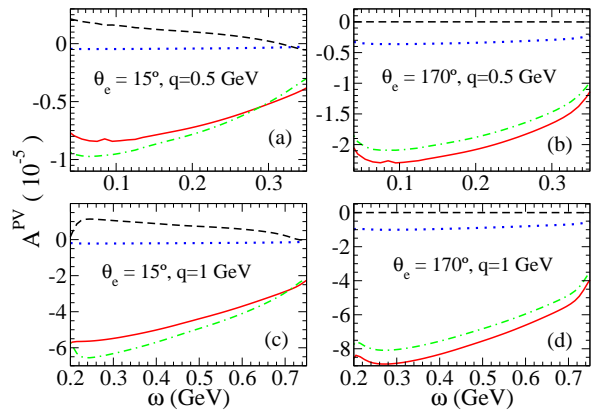}
    \caption{(Color online) Inclusive PVQE asymmetry (red line) and the three separate components: longitudinal (L, dashed black line), transverse (T, dashed-dotted green line) and axial-transverse (T', dotted blue line). 
    Results corresponding to two values of the momentum transferred $q=0.5$ GeV and $q=1$ GeV are shown in the upper and lower panels, respectively. 
    The forward (backward) scattering situation, $\theta_e=15^o$ ($\theta_e=140^o$), is represented in the panels on the left (right) side. 
    The NCC2 prescription is used to describe the current and the model RPWIA has been employed.}
    \label{fig:asi_contribuciones}
\end{figure}

To simplify the analysis of the results we decompose the PVQE asymmetry into a sum of three contributions:
\ba
{\cal A}^{PV}_{QE}={\cal A}_L + {\cal A}_T + {\cal A}_{T'} \, ,
\ea
where
${\cal A}_L$, ${\cal A}_T$ and ${\cal A}_{T'}$ are proportional to the corresponding PV responses:
$\widetilde{R}^L$, $\widetilde{R}^T$ and $\widetilde{R}^{T'}$.
%{\cal A}_L \propto \widetilde{R}^L$,
%{\cal A}_T \propto \widetilde{R}^T$ and,
%{\cal A}_{T'} \propto \widetilde{R}^{T'}$.
%
In Fig.~\ref{fig:asi_contribuciones} we present the total asymmetry and the three separated contributions as functions 
of the transferred energy, $\omega$.
As observed, the transverse term ${\cal A}_T$ dominates in all situations. Concerning the two remaining terms, the relative
predominance of one over the other depends on the specific kinematics:
${\cal A}_{L}$ dominates at forward scattering angles, whereas ${\cal A}_{T'}$ gets larger at backward angles.
Also, notice that ${\cal A}_{T'}$ is negligible at forward angles (likewise for ${\cal A}_{L}$ at backward angles).
These results apply to both $q$-values selected: $q=500$ MeV (upper panels) and $q=1000$ MeV (lower), and can be understood from the behavior of the leptonic factors, $v_L$, $v_T$ and $v_{T'}$, in addition to the values of the weak
coupling factors and the particular role played by the different nucleon form factors (see ~\cite{RaulThesis} for details).

\subsection{FSI and dynamical relativistic effects}
\label{FSIrelatdinam}

\begin{figure}[htbp]
    \centering
        \includegraphics[width=.34\textwidth,angle=270]{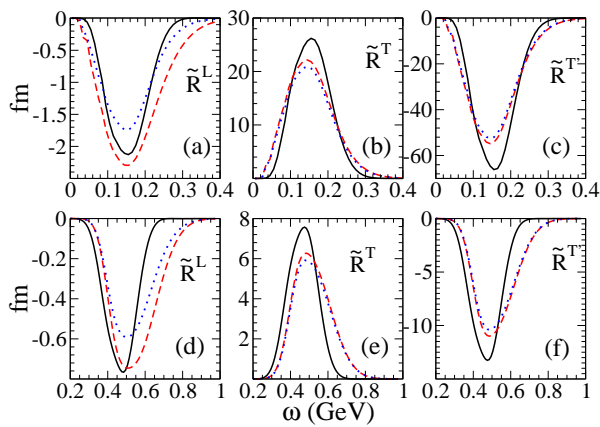}
    \caption{(Color online) PV responses computed with the models:
    RPWIA (black line), RMF-FSI (dashed red line) and EMA (dotted blue line). 
    In the upper (lower) panels the momentum transferred is $q=0.5$ GeV ($q=1$ GeV). 
    The prescription NCC2 has been used.}
    \label{fig:WNC_resp_models}
\end{figure}

\begin{figure}[htbp]
    \centering
        \includegraphics[width=.34\textwidth,angle=270]{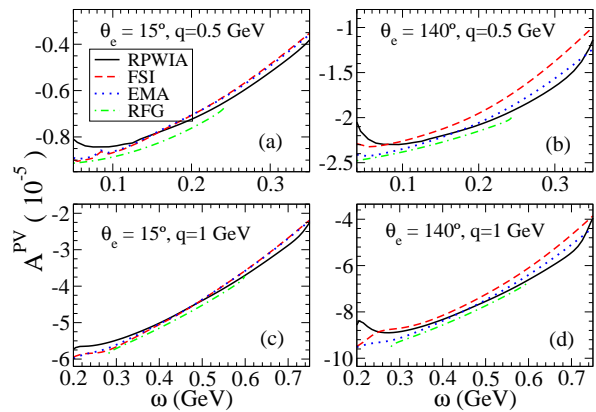}  
    \caption{(Color online) As in Fig.~\ref{fig:asi_contribuciones}, but in this case the PVQE asymmetry has been computed using the following models:
    RPWIA (solid black line), RMF-FSI (dashed red line), EMA (dotted blue line) and RFG (dashed-dotted green line).}
    \label{fig:asi_models}
\end{figure}

In Fig.~\ref{fig:WNC_resp_models} we present the PV responses computed
using the NCC2 prescription and the models: RPWIA and RMF-FSI.
Additionally, in order to estimate the effect of the lower components of the nucleon wave function we present the results 
computed in the so-called ``effective asymptotic momentum approximation''~\cite{Kelly97,Kelly99,Udias95,Udias01,Vignote04}.
Within this approach, which is simply denoted as EMA, the nucleon wave function is reconstructed by imposing that the 
relation between lower and upper components is the same as the one for free spinors, that is,
\ba
\psi_{d}({\bs p}) 
  = \frac{{\bs\sigma}\cdot{\bs p}_{as}}{E_{as}+M_N}\psi_u({\bs p}) \, , \label{EMAud}
\ea
where $\psi_{d}$ ($\psi_{u}$) represents the lower (upper) component of the nucleon wave function. 
The terms $E_{as}$ and ${\bs p}_{as}$ refer to the asymptotic energy and momentum of the nucleon, such that 
$E_{as}^2=M_N^2+p_{as}^2$. In the results shown in Figs.~\ref{fig:WNC_resp_models} and \ref{fig:asi_models}, the EMA 
model has been applied to both the bound and scattered nucleon wave functions; 
thus, for the scattered nucleon one has ${\bs p}_{as}\equiv{\bs p}_N$, whereas for the bound state 
${\bs p}_{as}\equiv{\bs p}_N-{\bs q}$.

As observed in Fig.~\ref{fig:WNC_resp_models}, the effects introduced by FSI (RMF-FSI {\it vs} RPWIA) are, on the one hand, 
a shift in the maximum of the responses to higher $\omega$-values, and, on the other, a significant change in their shape: a pronounced
asymmetry with more strength in the tail (high transferred energies).
Both effects tend to increase for larger $q$-values.
On the contrary, the projection on positive energy states (RMF-FSI {\it vs} EMA) does not change the shape of the responses, 
although it modifies the global magnitude. 

In what follows we discuss each response separately.
In the case of the transverse responses ($\widetilde{R}^T$ and $\widetilde{R}^{T'}$), FSI (RMF-FSI {\it vs} RPWIA) 
gives rise to a significant reduction of the maximum ($\sim20\%$).
On the other hand, the effect of projection over positive-energy states (EMA) also produces a slight decrease of the responses 
of the order of $6\%$.

The longitudinal response, $\widetilde{R}^L$, is approximately one order of magnitude smaller than the two other 
PV responses. Moreover, FSI produces a clearly different effect in this response. 
Contrary to the situation observed in the transverse channel, where FSI tends to decrease the maximum of the responses 
compared with the results in RPWIA, the global magnitude of $\widetilde{R}^L$ computed in RMF-FSI (dashed red line) is even larger than the result in RPWIA. 
On the contrary, the projected $L$ response (EMA, dotted blue line) is significantly reduced compared with the full FSI result. 
This behavior is clearly different from the one observed in the case of the 
longitudinal EM response (see, for instance, ~\cite{Caballero98a}) where its dependence with FSI and/or positive-energy projections follows a similar trend to the one shown by the transverse responses.
Finally, the smallness of $\widetilde{R}^L$ can be understood considering its dependence on the matrix 
elements of the current operator:
\ba
\widetilde{R}^L &\sim& \left(J_{EM}^{L,p}\right)^*J_{WNC,V}^{L,p}
    + \left(J_{EM}^{L,n}\right)^*J_{WNC,V}^{L,n}\nonumber\\
    &\approx& \left(J_{EM}^{L,p}\right)^*J_{WNC,V}^{L,p}
    + \left(J_{EM}^{L,n}\right)^*J_{WNC,V}^{L,p}\, , \nonumber \\
& & \label{eqeq}
\ea
where the label $n$ ($p$) refers to neutron (proton).
It is important to point out that the EM longitudinal response of the neutron is very small compared with 
the proton one due to the very minor contribution of the electric form factor of the neutron $G_E^n$; 
that is, $J_{EM}^{L,n} \ll J_{EM}^{L,p}$. 
Additionally, in the previous expression in Eq.~(\ref{eqeq}) it has been assumed that 
$G_{E}^p\approx-\widetilde{G}_{E}^n$. 
The smallness of the term $\widetilde{G}_E^pG_E^p$ makes in general
$J_{WNC,V}^{L,p}$ to be of the same order or even smaller than $J_{EM}^{L,n}$.
 
In what follows we investigate the impact that the previous ingredients have on the PVQE asymmetry. 
In Fig.~\ref{fig:asi_models} we present the asymmetry computed using the three models presented previously: 
RMF-FSI, EMA and RPWIA. We also add for reference the result corresponding to the relativistic Fermi gas (RFG) model 
(see ~\cite{Donnelly92,Musolf92} for details).

In the case of forward scattering (left panels), the maximum dispersion in the results is reached at the extreme $\omega$-values, 
{\it i.e.,} far from the quasielastic peak (QEP). RMF-FSI and EMA provide very similar asymmetries while RPWIA results 
differ, at most, by $\sim10\%$ at $\omega<0.1$ GeV and $q=0.5$ GeV.
It is important to point out that in the region around the QEP 
($\omega\sim 150$ MeV at $q=500$ MeV and $\omega\sim 500$ MeV at $q=1000$ MeV), 
the three models (RPWIA, EMA and RMF-FSI) lead to very similar results, with a dispersion lower than $\sim1\%$.

At backward scattering angles (right panels), the discrepancies between the three models get larger.
The highest dispersion is reached for $\omega$-values far from the center of the QEP. 
At $q=500$ MeV and $\omega\sim 50$ MeV the discrepancy is of the order of $15\%$ (RPWIA {\it vs} EMA). 
This difference holds at $\omega\sim 300$ MeV (RPWIA {\it vs} RMF-FSI). 
In the $\omega$-region close to the center of the QEP the dispersion between the curves is always smaller than $\sim 5\%$.

Finally, the RFG curves are the ones that deviate the most, particularly, from the results that incorporate FSI: EMA and RMF-FSI,
with the difference being somewhat smaller with respect to RPWIA results.
In the $\omega$-region where the responses reach their maxima (center of the QEP) the difference between RFG and RMF-FSI is 
lower than $7\%$. This difference diminishes for increasing values of the momentum transferred, $q$. 

\begin{figure}[htbp]
    \centering
        \includegraphics[width=.34\textwidth,angle=270]{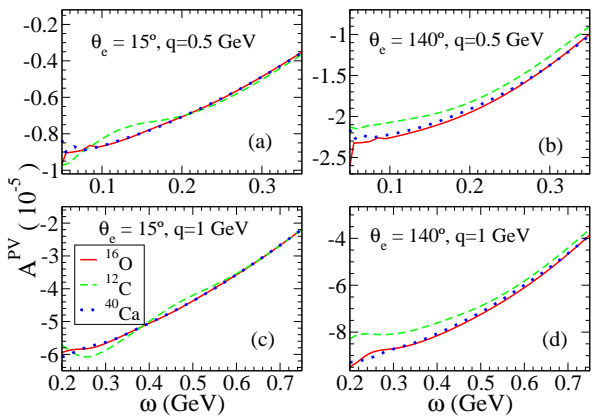}	  
    \caption{(Color online) PVQE asymmetry computed with RMF-FSI for several target nuclei:
    $^{16}$O (solid red line), $^{12}$C (dashed green line) and $^{40}$Ca (dotted blue line).
    Here the organization of the panels is the same as in Fig.~\ref{fig:asi_contribuciones}.}
    \label{fig:asi_nucleos}
\end{figure}

In the previous paragraph our interest has been placed on the role played by different descriptions of the final nucleon state. 
All results have been presented for the case of $^{16}$O as target nucleus.
In what follows we analyze the effects stemming from the use of different target nuclei. 
In Fig.~\ref{fig:asi_nucleos} we show the PVQE asymmetry corresponding to three different nuclear systems: 
$^{16}$O, $^{12}$C and $^{40}$Ca. All results have been computed within the RMF-FSI approach.
As observed, the largest differences are of the order of $10\%$ (at $\omega\sim 0.25$ GeV and $q=1$ GeV).
This result is consistent with previous studies~\cite{Amaro96} and it proves that ${\cal A}^{PV}_{QE}$ can provide a useful tool, 
complementary to the elastic electron-proton asymmetry, to get accurate information on the electroweak structure of the nucleon.
This subject is discussed at length in the next sections.

\subsection{Off-shell effects}\label{offshell}

Our aim in this section is to investigate the role of off-shell effects in the PVQE asymmetry. Contrary to the case of 
elastic electron-proton scattering, the use of nuclear targets requires the description of nucleons that are not on-shell. 
This subject has been
treated in detail within the framework of coincidence $(e,e'N)$ reactions. Moreover, its study in the case of PV
electron scattering has been presented in an accompanying paper~\cite{exclusive}. Here we extend these investigations to the
case of inclusive electron scattering. As is well known, the off-shell character of the nucleons involved in electron-nucleus scattering
leads to results, cross sections and nuclear response functions, that can differ significantly when different expressions 
for the nucleon current operator and/or different gauges are selected. 

In this section we follow the general terminology introduced in our previous work 
(see ~\cite{Caballero98a,Caballero98b,Martinez02b,Caballero93} for details) and analyze how the PVQE asymmetry is modified when several prescriptions are used: 
NCC1, NCC2, CC1(0), CC2(0), CC1(3) and CC2(3).
All results in this section have been computed using the RMF-FSI model, that is, FSI are incorporated in the general formalism making
use of the same relativistic, scalar and vector, potentials used for the bound nucleon states. This model has been widely tested
in the case of the purely EM interaction comparing its prediction with a large body of $(e,e')$ data 
(see ~\cite{Caballero06,Meucci09,Gonzalez-Jimenez14b}).

\begin{figure}[htbp]
    \centering
        \includegraphics[width=.345\textwidth,angle=270]{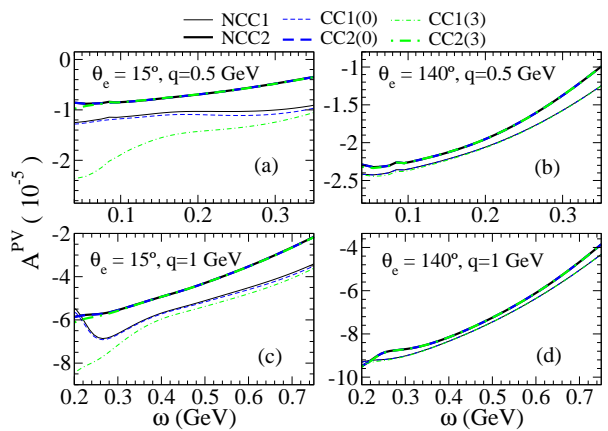}
    \caption{(Color online) PVQE asymmetry within RMF-FSI model, with the
    same organization of the panels as in Fig.~\ref{fig:asi_contribuciones}.
    The results are computed using CC1 and CC2 currents in the different gauges (see legend).}
    \label{fig:asi_CC1vsCC2_rmf}
\end{figure}

In Fig.~\ref{fig:asi_CC1vsCC2_rmf} we analyze the sensitivity of the PVQE asymmetry against different off-shell prescriptions.
We observe that ${\cal A}_{QE}^{PV}$ shows a tiny dependence with the gauge selected except in the case of CC1(3) at 
forward scattering angles (left panels). Although not presented here, the longitudinal responses
from CC1(3) are shown to be dramatically different from the corresponding responses associated with the remaining prescriptions.
Moreover, the choice of the nucleon current operator (CC1 {\it vs} CC2) leads in general to important differences.
At forward angles (left panels) the discrepancies are of the order of $\sim 30\%$ ($\sim 17\%$) at $q=0.5$ GeV ($q=1$ GeV) 
in the $\omega$-region close to the center of QEP. On the contrary, at backward angles (right panels) the differences 
are considerably reduced: $\sim 5\%$ ($\sim 2.5\%$) at $q=0.5$ GeV ($q=1$ GeV) in the same $\omega$-region.

\subsection{Nucleon structure: WNC form factors}\label{efecnucleonicos}

In this section we evaluate the impact that the description of the nucleon form factors has in the PVQE asymmetry ${\cal A}_{QE}^{PV}$. 
One of the ingredients that makes the study of this observable appealing is the presence of the neutronic channel in the scattering process.
In order to highlight the differences with the elastic case, one can separate the contributions from protons and neutrons in the 
PVQE asymmetry, {\it i.e.,} 
${\cal A}^{PV}_{QE} = {\cal A}_p + {\cal A}_n$, where
\ba
{\cal A}_{p,n} &=& \frac{{\cal A}_0}{2{\cal G}^2} 
\left[a_A\left(v_L\widetilde{R}_{p,n}^L +v_T\widetilde{R}_{p,n}^T\right)\right.\nonumber\\
         &-& \left.a_Vv_{T'}\widetilde{R}_{p,n}^{T'}\right]\,,\label{APVQEpn}
\ea
Where $\widetilde{R}_{p,n}$ are the proton ($p$) and neutron ($n$) PV responses.
The analysis presented in this work applies only to the case of zero isospin nuclei, {\it i.e.,} nuclei with the same number of protons and neutrons. 
Although the results presented in this section correspond to $^{16}$O, we have checked that the discussion follows similar trends
in the case of other nuclei such as $^{12}$C and $^{40}$Ca.

In this work we use the GKex prescription to describe 
the behavior of $G_{E,M}^{p,n}$, whereas for the WNC form factors we make use of the results obtained in 
~\cite{Gonzalez-Jimenez13a}, also briefly summarized in Sect.~\ref{WNCff}.

Before entering into a detailed discussion of the results, some comments on the general procedure considered are in order.
We have checked that $\widetilde{R}^T$, computed within the RMF-FSI approach, does show a small dependence on the WNC electric 
form factor $\widetilde{G}_E$; therefore, the asymmetry contribution ${\cal A}_T$ is very insensitive to the electric strangeness content.
Similarly, $\widetilde{R}^L$ shows a weak dependence on $\widetilde{G}_M$;
then the magnetic strange form factor $G_M^s$ plays a very small role in the asymmetry ${\cal A}_L$. 
Finally, the PV response $\widetilde{R}^{T'}$ (see ~\cite{exclusive}) is, by construction, 
independent of $\widetilde{G}_M$ and $\widetilde{G}_E$; 
hence neither the electric nor the magnetic strangeness can modify ${\cal A}_{T'}$. Therefore, the dependence of the PV asymmetry with the nucleon strangeness 
enters only through the $T$ channel (magnetic strangeness) and the $L$ one (electric). 
The latter only occurs at very forward scattering angles. In what follows we discuss
these results in detail.
%

%%%%%%%%%%%% EXTRA\~NEZA MAGNETICA %%%%%%%%%%%%%%%%%%%%%

\begin{figure}[htbp]
    \centering
        \includegraphics[width=.23\textwidth,angle=270]{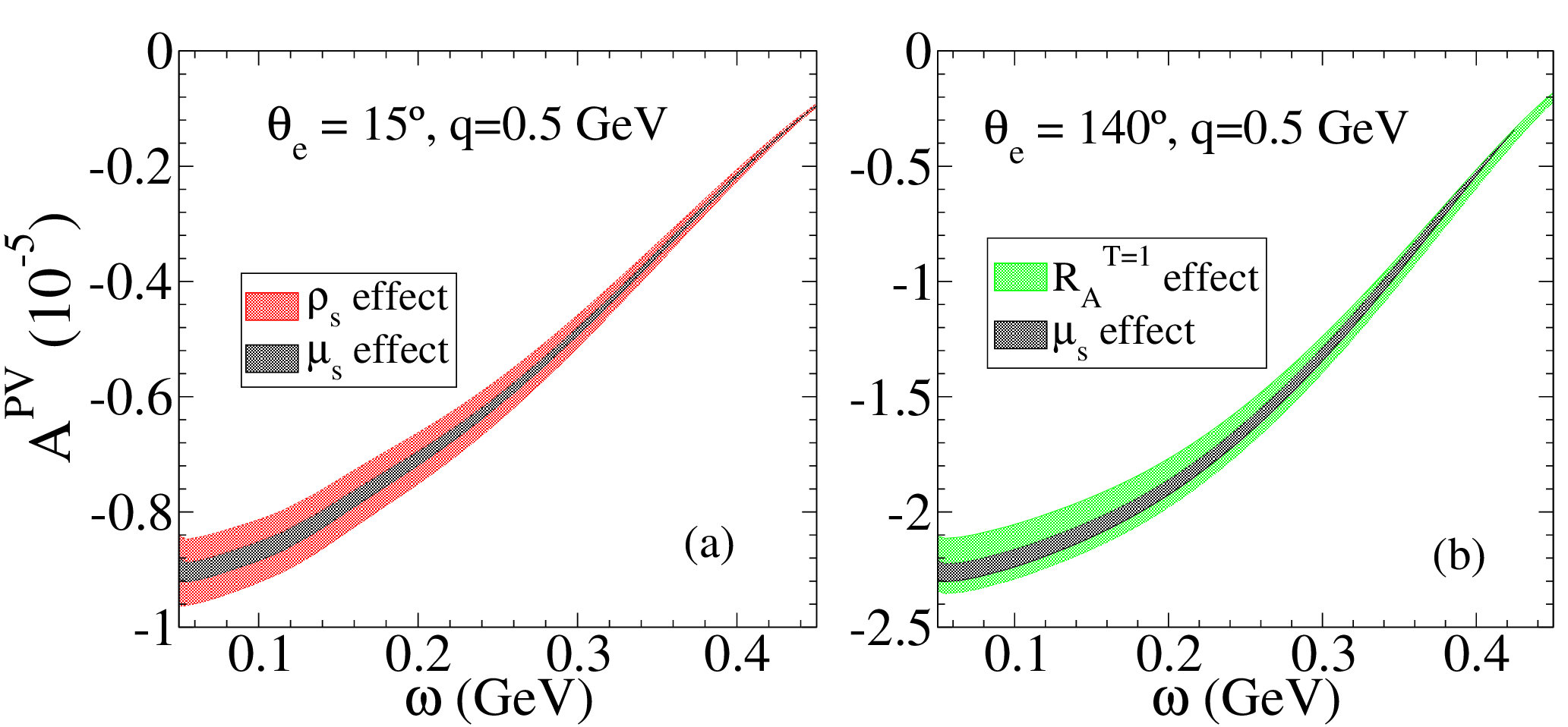}\\
        \includegraphics[width=.23\textwidth,angle=270]{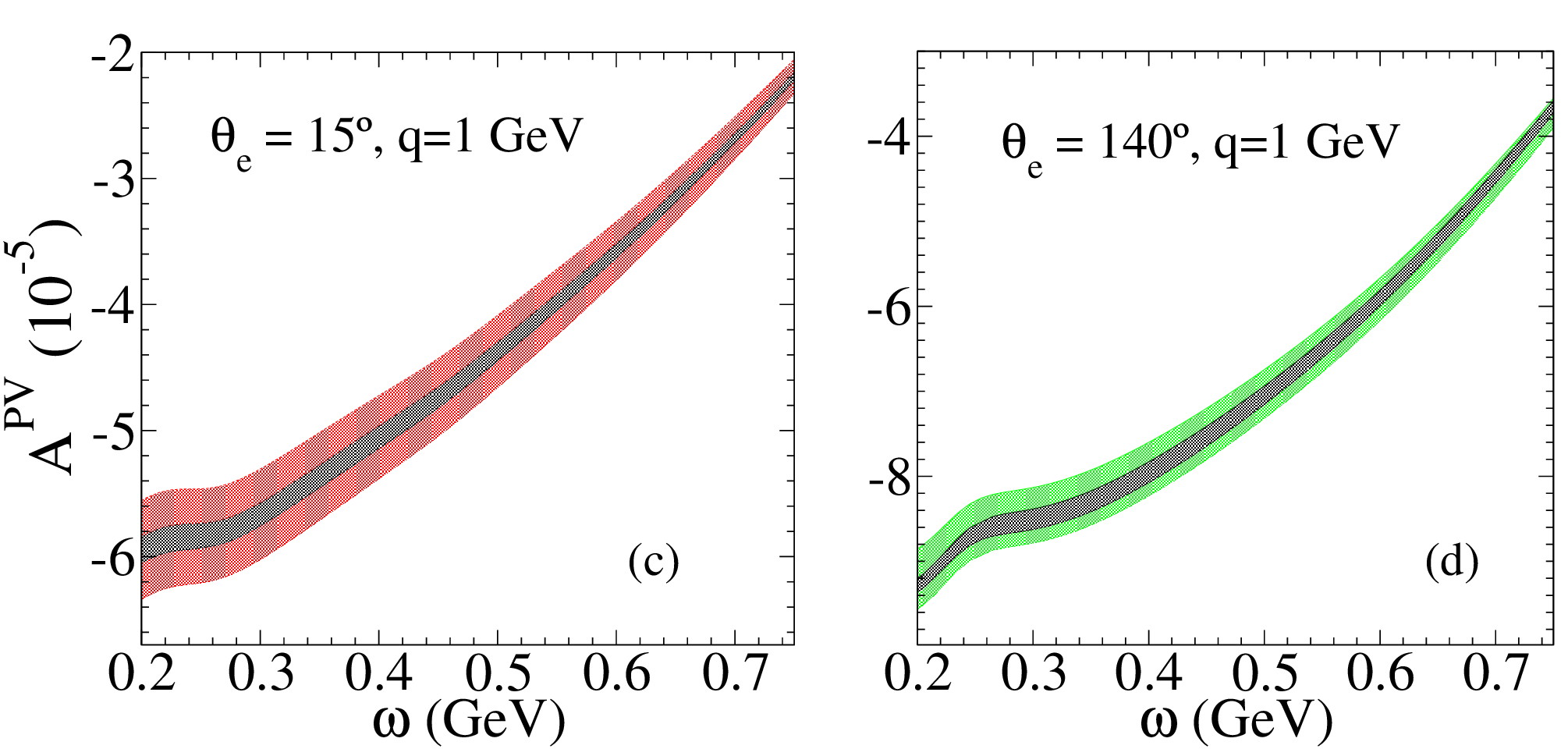}
    \caption{(Color online) PVQE asymmetry at forward (left panels) and backward (right panels) scattering angles. 
    The upper (lower) panels correspond to the momentum transfer $q=0.5$ GeV  ($q=1$ GeV).
    The black band represents the uncertainty in ${\cal A}^{PV}_{QE}$ due to the magnetic strange parameter $\mu_s$.
    The red band (left panels) shows the effect of the electric strange parameter $\rho_s$ while the green band (right panels) 
    corresponds to the impact associated with the uncertainty in the isovector contribution to the axial-vector form factor (see text for details).}
    \label{fig:asi_nucleonic}
\end{figure}

To address the impact on ${\cal A}_{QE}^{PV}$ linked to the description of the magnetic strange form factor, we have computed the PVQE asymmetry using the two extreme values of $\mu_s=-0.02\pm0.21$. 
This is represented in Fig.~\ref{fig:asi_nucleonic} by the black band.
We observe that at forward scattering angles (left panels) 
the width of the band is: $\sim4\%$ ($\sim3.5\%$) at $q=500$ MeV ($q=1000$ MeV). 
Similar results are found at the backward kinematics (right panels):
$\sim3.5\%$ for both $q$-values, $q=500$ and $1000$ MeV.

This low sensitivity of ${\cal A}^{PV}_{QE}$ to the magnetic strange form factor can be easily understood.
As a first approximation, the PV transverse response $\widetilde{R}^T$ can be simply given by the particular combination: 
$\widetilde{R}^{T}_{p,n}\sim G_M^{p,n}\widetilde{G}_M^{p,n}$. Hence we can write: 
\ba
\widetilde{R}^T_p \sim (1-4\sin^2\theta_W)(1+R_V^p)(G_M^p)^2\nonumber\\
       - (1+R_V^n)G_M^pG_M^n + (1+R_V^{(0)})G_M^pG_M^s\,,
\ea
and
\ba
\widetilde{R}^T_n \sim (1-4\sin^2\theta_W)(1+R_V^n)(G_M^n)^2\nonumber\\
       - (1+R_V^p)G_M^nG_M^p + (1+R_V^{(0)})G_M^nG_M^s\,.
\ea
Assuming $G_M^p\approx-G_M^n$, it can be seen that the nucleon magnetic strangeness does play a very minor role in the transverse response,
no matter which specific scattering angle is considered. 
In other words, when adding proton and neutron contributions, the last term goes as the isoscalar magnetic form factor, which is very much 
smaller than the isovector one. 
Note that in a case like $^{27}$Al there will not be as good a cancellation, which may be interesting 
for the Qweak experiment~\cite{Qweak13} where some of the PV asymmetry comes from the aluminum windows.
In summary, the nucleon magnetic strangeness is also strongly reduced in the PVQE asymmetry, being
much smaller than the one found in the case of elastic electron-proton scattering~\cite{Gonzalez-Jimenez13a}. 

%%%%%%%%%%%% EXTRAÑENEZA ELECTRICA %%%%%%%%%%%%%%%%%%%%%

To study the effect of the electric strangeness in ${\cal A}_{QE}^{PV}$ we restrict ourselves to the forward scattering kinematics where 
the longitudinal contribution attains its largest value (see Fig.~\ref{fig:asi_contribuciones}).
We consider the two extreme values for the electric strange parameter: $\rho_s=0.59\pm0.62$.
This produces a dispersion of the order of $\sim13\%$ in the PVQE asymmetry (red band in left panels in Fig.~\ref{fig:asi_nucleonic}). 

The longitudinal PV response can be approximated as $\widetilde{R}^{L}_{p,n}\sim G_E^{p,n}\widetilde{G}_E^{p,n}$. 
Thus one can write:
\ba
\widetilde{R}^L_p \sim (1-4\sin^2\theta_W)(1+R_V^p)(G_E^p)^2\nonumber\\ 
      - (1+R_V^n)G_E^pG_E^n + (1+R_V^{(0)})G_E^pG_E^s\,,
\ea
and
%%%
\ba
\widetilde{R}^L_n \sim (1-4\sin^2\theta_W)(1+R_V^n)(G_E^n)^2\nonumber\\
      - (1+R_V^p)G_E^nG_E^p + (1+R_V^{(0)})G_E^nG_E^s\,.
\ea
Because of $G_E^n \ll G_E^p$, the role played by the electric strangeness is much weaker in $\widetilde{R}^L_n$ than in $\widetilde{R}^L_p$.
This means that the impact of $G_E^s$ in the PVQE asymmetry comes almost exclusively from the proton response.

From the previous discussion, a clear difference emerges between the present QE regime and the elastic one described
in \cite{Gonzalez-Jimenez13a}. Due to the minor role played by the magnetic strangeness in the PVQE asymmetry ($<4\%$), 
the measurement of ${\cal A}^{PV}_{QE}$ at 
forward kinematics could provide valuable information on $\rho_s$, being rather independent of $\mu_s$. This result is clearly
in contrast with the situation that is observed for the PVep asymmetry where $\rho_s$ and $\mu_s$ are strongly correlated~\cite{Gonzalez-Jimenez14a}. 
In this sense, the analysis of the QE regime could help in getting additional information
on the electric strangeness content in the nucleon, {\it i.e.,} $\rho_s$ (or $G_E^s$).
However, some caution should be drawn before arriving at definite conclusions. The analysis of the forward scattering situation 
is not free from ambiguities. We have already shown that off-shell effects may introduce significant uncertainties in the PVQE asymmetry.

%%%%%%%%%%%%% RC in the axial form factor %%%%%%%%%%%%%%%%%%%%%

To conclude, we analyze the sensitivity of the PVQE asymmetry with the axial-vector form factor. 
The axial transverse PV response can be approximated by the product of the magnetic and axial form factors,
{\it i.e.,} $R^{T'}_{p,n}\sim G_M^{p,n}G_A^{e,(p,n)}$.
Then, the following expressions hold:
\ba
\widetilde{R}^{T'}_p \sim -2(1+R_A^{T=1})G_A^{T=1}G_M^p &&\nonumber\\
          + \sqrt{3}R_A^{T=0}G_A^{(8)}G_M^p + (1+R_A^{(0)})G_A^sG_M^p\,,
\ea
and
\ba
\widetilde{R}^{T'}_n \sim 2(1+R_A^{T=1})G_A^{T=1}G_M^n\nonumber\\
          + \sqrt{3}R_A^{T=0}G_A^{(8)}G_M^n + (1+R_A^{(0)})G_A^sG_M^n\,.
\ea
As already shown for the purely transverse response $\widetilde{R}^T$, the approximation $G_M^p\approx-G_M^n$ leads to similar (but opposed) 
proton and neutron contributions to the PVQE asymmetry. 
Because of that, the effect of the axial-vector strangeness and the contribution 
from the octet isoscalar $G_A^{(8)}$ are very small. 
On the contrary, this response shows a strong sensitivity against any 
variation in the isovector contribution of the axial-vector form factor. 
This analysis is presented in the right panels of Fig.~\ref{fig:asi_nucleonic}. 
The ambiguity associated to the use of the extreme values: 
$R_A^{(T=1)}=0.082$ and $R_A^{(T=1)}=-0.598$, is of the order of $\sim10\%$ at $q=500$ MeV, while at $q=1000$ MeV it is slightly lower, 
$\sim8\%$. This is represented by the green band in the right
panels of Fig.~\ref{fig:asi_nucleonic}.

The entire analysis presented in this work corresponds to the impulse approximation.
Effects linked to meson-exchange currents (MEC),
only partially treated at present for PV electron scattering reactions, can also introduce differences in ${\cal A}_{QE}^{PV}$ 
at forward kinematics (see ~\cite{Amaro02}). 
Hence some caution should be addressed before more definite conclusions can be drawn.

\section{Scaling in the PV responses}

%%%%%%%%%%% Introduction %%%%%%%%%%%%

The analysis of inclusive electron scattering data in the QE domain has proven the validity of the scaling phenomenon. This means
that the reaction mechanism in the process can be properly described as the scattering between the electron and the constituents,
the nucleons, in the nuclear target. Hence the differential $(e,e')$ cross section divided by an appropriate single-nucleon cross section leads
to the so-called scaling function that is shown to depend only on a single variable, named the scaling variable, assuming the transferred momentum
is high enough. Moreover, this function scales with the nuclear species as the inverse of the Fermi momentum. Hence an universal 
superscaling function, namely, independent of the transferred momentum and the nuclear system, can be introduced. This property has been
shown to be fulfilled quite well by the longitudinal $(e,e')$ data, while it is violated in the transverse channel, where ingredients
beyond the impulse approximation come into play: $\Delta$-resonance, meson exchange currents, multi-nucleon excitations, {\it etc.} 

The superscaling approach has been studied in detail in the past~\cite{Day90,Alberico88,Donnelly99a,Donnelly99b,Maieron02,Meucci09} and its predictions have been extended to the analysis of neutrino-nucleus scattering processes~\cite{Amaro05a,Amaro05b,Caballero05,Caballero06,Amaro06,Caballero07,Gonzalez-Jimenez13b,Gonzalez-Jimenez13c,Megias13,Gonzalez-Jimenez14b,Megias14}. 
Moreover, scaling properties have also been analyzed within the context of different models, in particular,
the RPWIA and the RMF-FSI approaches considered in this work. 
One of the most outstanding results concerns the behavior shown by the superscaling
function extracted from the RMF-FSI model. Not only does the function show an important asymmetry, with a long tail extended to high transferred
energies, in accordance with data, but also the scaling functions corresponding to the longitudinal and transverse channels present differences,
the $T$ response being larger by $\sim 20\%$. 
This result, that is absent in most other theoretical approaches, seems to be supported by the analysis of data presented in ~\cite{Maieron09,Gonzalez-Jimenez14b}. 

In this work we extend for the first time the scaling/superscaling analysis to the PV responses. We apply this
study to our models and evaluate the interference scaling functions, $\widetilde{f}_{L,T,T'}$, by dividing the corresponding
PV nuclear responses by the appropriate single-nucleon contributions. The explicit expressions for the latter are given in
the Appendix~\ref{appScaling}. In order to make clear how scaling arguments work for the PV interference observables, these are
compared with the purely EM responses as well as with data.

%%%%%%%%% RESULTs %%%%%%%%%%%%%%%%

\begin{figure}[htbp]
    \centering
        \includegraphics[width=.40\textwidth,angle=0]{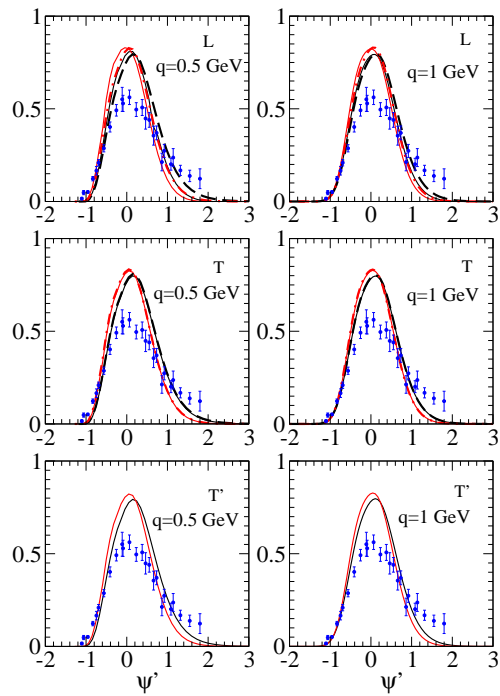}
    \caption{(Color online) EM (dashed lines) and interference (solid lines) scaling functions computed using NCC2 prescription, the RPWIA model and two target nuclei:
    $^{12}$C (black) and $^{16}$O (red).
    The left (right) panels correspond to the momentum transfer $q=500$ MeV ($q=1000$ MeV).
    We represent the longitudinal ($L$, upper panels), transverse ($T$, central panels) and transverse axial ($T'$, lower panels) scaling functions.
    As reference, the experimental longitudinal scaling data are also represented (blue points)~\cite{Maieron02}.}
    \label{fig:scaling_rpwia_cc2_nucleos}
\end{figure}

\begin{figure}[htbp]
    \centering
        \includegraphics[width=.39\textwidth,angle=0]{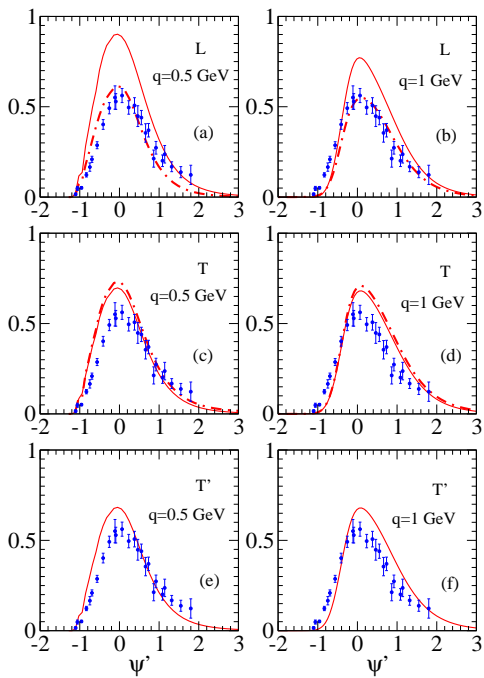}
    \caption{(Color online) As in Fig.~\ref{fig:scaling_rpwia_cc2_nucleos}, but using the RMF-FSI model and only for $^{16}$O.}
    \label{fig:scaling_rmf_cc2_nucleos}
\end{figure}

In Fig.~\ref{fig:scaling_rpwia_cc2_nucleos} we present the EM and PV interference (referred simply as PV) scaling functions computed with 
the NCC2 prescription and corresponding to the RPWIA model. Results are shown for two different target nuclei: $^{12}$C and $^{16}$O. 
Although not shown, results are very similar for $^{40}$Ca.  
Two values of the transferred momentum have been selected: $q=500$ MeV/c (left panels) and $q=1000$ MeV/c (right panels). In both
cases the scaling function is presented for the three (two) channels involved in PV (EM) electron scattering processes: $L$ (upper
panels), $T$ (medium) and $T'$ (lower). In all cases we include for reference the data corresponding to the analysis of the purely
EM longitudinal channel, {\it i.e.,} $f_L^{exp}$.

As observed, RPWIA fulfills first, second and third kinds of scaling, namely, the functions are almost independent of the transferred
momentum, the nuclear system and the particular channel considered. 
Moreover, the new PV scaling functions coincide with the purely EM ones.
This supports the idea of an universal scaling function. 
However, RPWIA theoretical predictions do not reproduce the shape nor the height of $f_L^{exp}$. 
As already discussed in previous work~\cite{Gonzalez-Jimenez14b}, the RPWIA leads to scaling functions that lack the strong asymmetry shown by the analysis of data. 
This behavior also applies to the PV responses.

In Fig.~\ref{fig:scaling_rmf_cc2_nucleos} we present the scaling functions but with FSI described by means of the RMF model. This analysis has
been shown in detail in the past for the EM responses. Here we extend our investigation to the PV observables. To make clearer the discussion we
only show results for $^{16}$O. Other nuclear systems, such as $^{12}$C and $^{40}$Ca, lead to the same conclusions with very similar results.
Contrary to the RPWIA case, note that the RMF-FSI model leads to scaling functions that show some dependence with $q$, {\it i.e.,} scaling
of the first kind is broken at some degree. However, this is consistent with PC electron scattering data and their associated uncertainties. In particular,
for increasing $q$-values (compare left panels, $q=0.5$ GeV, with the right ones, $q=1$ GeV), the peak of the scaling function is shifted
to higher $\psi$ (higher $\omega$ values), the maximum being reduced and the tail enhanced. Concerning the comparison
with data (strictly speaking this should only apply to the purely EM longitudinal channel), the accordance improves very significantly with
respect to the RPWIA predictions; RMF-FSI is able to reproduce not only the height of the peak, but also the particular asymmetrical shape
of $f_L^{exp}$. 

With regards to the comparison between the EM and the interference scaling functions evaluated with the RMF-FSI model 
(Fig.~\ref{fig:scaling_rmf_cc2_nucleos}), in the transverse channel one gets
$f_T\approx\widetilde{f}_T \approx\widetilde{f}_{T'}$. 
On the contrary, the role played by FSI is clearly different in the longitudinal channel. Notice that the weak
interference longitudinal scaling function, $\widetilde{f}_L$, is significantly larger than the purely EM one, $f_L$:
$f_L\approx0.6\widetilde{f}_L$ ($f_L\approx0.75\widetilde{f}_L$) at $q=500$ MeV (1000 MeV). 
It is important to point out that RMF-FSI leads to a function $\widetilde{f}_L$ with its maximum being considerably higher than the
corresponding result in RPWIA. This behavior is in contrast to the effects introduced by FSI for all the remaining scaling functions,
including the EM longitudinal one. This particular result can be connected with the smallness of the PV longitudinal
response (see Fig.~\ref{fig:WNC_resp_models}), that consequently shows a very high sensitivity to the distortion introduced by FSI.
Notice that $f_L\approx\widetilde{f}_L$ within RPWIA.

To conclude, scaling of zeroth kind is clearly violated by both the EM and the PV functions. In the latter,
one can express: $\widetilde{f}_T\approx0.75\widetilde{f}_L$ ($\widetilde{f}_T\approx0.95\widetilde{f}_L$) at $q=500$ MeV (1000 MeV), that is,
the longitudinal function exceeds the transverse contribution. 
In contrast, the EM responses satisfy: 
$f_L\approx0.85f_T$ at both $q$-values. 
This result is consistent with previous studies~\cite{Gonzalez-Jimenez14b} and with the preliminary analysis of the separated EM transverse $(e,e')$ data performed by Donnelly and Williamson~\cite{DonnellyWilliamson} (see also ~\cite{Maieron09}).
Although not shown, similar results are obtained for $^{12}$C and $^{40}$Ca. 
Hence scaling of second kind, namely, independence on the nuclear system, is fulfilled within the RMF-FSI model.

\section{Summary and Conclusions}

This work has been devoted to the study of PVQE electron-nucleus scattering. 
Our main interest has been to explore new observables (in addition to the ones occurring for the elastic electron-nucleon reaction) 
that allow us to get new and precise information on the nucleon structure. In particular, PVQE reactions on complex nuclei can provide 
information on the WNC form factors that complements the one obtained from other processes such as elastic scattering 
off protons~\cite{Gonzalez-Jimenez13a,Gonzalez-Jimenez14a} and light nuclei~\cite{HAPPEXa,HAPPEXHe,Moreno13,Moreno14}, QE 
electron scattering off deuterium~\cite{SAMPLE05,G010}, neutrino scattering, {\it etc.}

To simplify the analysis we have isolated in the PV asymmetry (${\cal A}^{PV}_{QE}$) its $L$ contribution (linked to 
$\widetilde{R}^L$), $T$ (linked to $\widetilde{R}^T$) and $T'$ (linked to $\widetilde{R}^{T'}$).
We have found that the $T$ component dominates for all kinematics; on the contrary, 
the $L$ contribution is negligible at backward scattering angles, while $T'$ is small at forward angles.

We have performed a fully relativistic description of PV $(\vec{e},e')$ processes and have quantified the uncertainties in the responses and 
the asymmetry linked to the following ingredients:
\begin{itemize}
 \item {\it Treatment of FSI and description of the nucleon wave function (Sect.~\ref{FSIrelatdinam})}.
 
The addition of FSI produces a visible change in the shape of the responses (symmetric in RPWIA and asymmetric with a more pronounced tail with FSI on), 
that are also shifted to higher $\omega$-values.
In particular, FSI reduce the height of $R^{L,T}$ and $\widetilde{R}^{T,T'}$ (in their maxima) around $15-20\%$.
The role played by the distortion of the lower components of the nucleon wave functions (FSI {\it vs} EMA) in the PV responses
is of the order of $\sim5-10\%$. However, a different behavior is shown for the PV longitudinal response $\widetilde{R}^L$. 
Here FSI do not reduce the height of the maximum, but tend to increase the total area under the response, which is notably larger than in RPWIA.
This result can be a consequence of the smallness of $\widetilde{R}^L$ (one order of magnitude smaller than the others). 
Dynamical relativistic effects, {\it i.e.,} distortion of the lower components in the nucleon wave functions, make a very significant difference
in $\widetilde{R}^L$. Notice that $\widetilde{R}^L$ computed with EMA presents a behavior similar to the rest of responses. 
This is in contrast to $\widetilde{R}^L$ evaluated with full FSI.

With regards to the PVQE asymmetry, we find that in the region of $\omega$ close to the maximum of the QE peak 
the difference between RPWIA, FSI-RMF and EMA results is $\sim1\%$ ($\sim5\%$) at forward (backward) scattering angles. 
For values of $\omega$ away from the center of the QE peak those differences are always below $\sim10\%$ ($\sim15\%$) 
at forward (backward) angles.

\item {\it Description of the hadronic vertex (off-shell effects, Sect.~\ref{offshell})}.
 
Effects in the transverse responses $R^T$ and $\widetilde{R}^{T,T'}$ (with FSI) linked to the choice of the current operator 
(CC1 {\it vs} CC2) deserve to be commented on: differences are of the order of $20-40\%$. These discrepancies show up in the PVQE asymmetry. 
At forward scattering angles the effects are of the order of $15-30\%$ in the region of $\omega$ around the maximum of the responses. 
At backward angles the differences are reduced, notably being lower than $5\%$ in the same region of $\omega$. 
Finally, the three gauges, Landau, Coulomb and Weyl, provide very similar responses when using CC2 and RMF-FSI.
On the contrary, the use of CC1 leads to significant differences in the responses, particularly, in the case of the Weyl gauge. 
The effects in the asymmetry due to the choice of the gauge are tiny at backward scattering angles because of the negligible contribution 
of the longitudinal responses. These effects remain small at forward scattering angles except for the CC1(3) prescription.
\end{itemize}

In addition to the effects associated with the nuclear model description, in Sect.~\ref{efecnucleonicos} the 
sensitivity of the PVQE asymmetry to the nucleon form factors has been investigated. The PV asymmetry shows a very mild dependence 
with the magnetic strangeness content in the nucleon because of cancellations between the proton and neutron contributions. 
A similar comment applies to the isoscalar contributions (including the axial-vector strangeness).
Regarding the sensitivity of the PVQE asymmetry on the electric strangeness content, at forward scattering angles it is of the 
order of $\sim13\%$ (for the $q$-values considered in this work). This result has been estimated by using the extreme values of the 
parameter $\rho_s=0.59\pm0.62$ that are consistent with the analysis of the PVep asymmetry data presented in 
~\cite{Gonzalez-Jimenez13a,Gonzalez-Jimenez14a}. It is important to point out that getting nucleonic information from 
measurements of ${\cal A}^{PV}_{QE}$ at forward scattering angles is not free from ambiguities. 
On one hand, the choice of CC1 and/or CC2 (off-shell effects) gives rise to differences in the asymmetry of the order of $\sim30\%$ ($\sim17\%$) 
at $q=500$ MeV ($q=1000$ MeV). On the other hand, effects linked to MEC could modify in a significant way the 
results obtained at these kinematics (see ~\cite{Amaro02}).

Choosing backward scattering kinematics makes the analysis of results much more favorable. Here the choice of $R_A^{T=1}$ within the range given 
by $[-0.598\, , \, 0.082]$ produces a change in the PVQE asymmetry of the order of $\sim10\%$ ($\sim8\%$) at $q=500$ MeV ($q=1000$ MeV).
At backward angles MEC effects in the asymmetry are expected to be small: below $\sim0.5\%$ at $q=500$ MeV and much smaller 
at higher $q$ (see ~\cite{Amaro02}), and furthermore, off-shell effects are also significantly reduced: 
$\sim5\%$ ($\sim2.5\%$) at $q=500$ MeV ($q=1000$ MeV). Therefore, a measurement of the PVQE asymmetry at backward scattering angles 
and transferred momentum $q\sim 500-1000$ MeV could be very useful to estimate the radiative corrections that enter in the isovector 
axial-vector sector of the weak neutral current, $R_A^{T=1}$. 

An important effect in the determination of the strange form factors comes from the uncertainty linked to $R_A^{T=1}$ due to the correlation 
between $\mu_s$ and $R_A^{T=1}$ at backward scattering angles. 
In other words, $\mu_s$ values obtained from the analysis of the PVep asymmetry data~\cite{Gonzalez-Jimenez13a,Liu07} are affected by the value of $R_A^{T=1}$ employed in the fit. 
Moreover, due to the strong correlation between $\mu_s$ and $\rho_s$, 
the value of the latter depends strongly on the value of the former. 
This is clearly illustrated in ~\cite{Gonzalez-Jimenez14a}.
Thus, a more accurate determination of $R_A^{T=1}$ would reduce 
significantly the theoretical uncertainties associated with the vector strange form factors. This would also establish constraints that 
any theoretical model aiming to describe the so-called anapole effects (implicit in $R_A^{T=1}$) should fulfill.

% \vspace{0.25cm}

\section*{Acknowledgements}

This work was partially supported by DGI (Spain): FIS2011-28738-C02-01, 
by the Junta de Andaluc\'ia (FQM-160) and
the Spanish Consolider-Ingenio 2000 programmed CPAN, 
and in part (TWD) by US Department of Energy under grant Contract Number
DE-FG02-94ER40818.
R.G.J. acknowledges financial help from VPPI-US (Universidad de Sevilla) 
and from the Interuniversity Attraction Poles Programme initiated 
by the Belgian Science Policy Office.

%%%%%%%%%%%%%%%%%%%%%%%
%%%%%%%%%%%%%%%%%%%%%%%
%%%%%%%%%%%%%%%%%%%%%%%
%%%%%%%%%%%%%%%%%%%%%%%
\appendix

\section{Definitions of the interference scaling functions}\label{appScaling}

In the context of the relativistic Fermi gas (RFG) the scaling variable is defined as (see ~\cite{Alberico88,Donnelly99a,Donnelly99b})
\ba
\psi' \equiv \frac{1}{\sqrt{\xi_F}}\frac{\lambda'-\tau'}
             {\sqrt{(1+\lambda')\tau'+\kappa\sqrt{\tau'(\tau'+1)}}}\,,
\ea
where $\xi_F = \sqrt{1+(k_F/M)^2}-1$, $\kappa=q/(2M)$, $\lambda'=\omega'/(2M)$ 
and $\tau=\kappa^2-\lambda'^2$. 
$M$ is the nucleon mass and $k_F$ the Fermi momentum~\cite{Maieron02}. 
We have introduced the variable $\omega'$ defined as $\omega'=\omega-E_{shift}$, where, the quantity $E_{shift}$ depend on the target nucleus~\cite{Maieron02}. 

The EM longitudinal, $L$, and transverse, $T$, scaling functions are defined in ~\cite{Donnelly99b}.
% \ba
% f_{L,T} \equiv k_F \frac{R_{L,T}(\kappa,\lambda)}
%                        {G_{L,T}(\kappa,\lambda)}\,;
% \ea
Similarly, the interference scaling functions are:
\ba
\widetilde{f}_{L,T,T'} \equiv k_F \frac{\widetilde{R}_{L,T,T'}(\kappa,\lambda)}
                       {\widetilde{G}_{L,T,T'}(\kappa,\lambda)}\,.
\ea
We have introduced the functions:
\ba
% G_{L,T}(\kappa,\lambda)
%      =\frac{1}{2\kappa {\cal D}}U_{L,T}(\kappa,\lambda)\,,\\
%%%%%%
\widetilde{G}_{L,T,T'}(\kappa,\lambda)
     =\frac{1}{2\kappa {\cal D}}\widetilde{U}_{L,T,T'}(\kappa,\lambda) \, ,
\ea
where
\ba
% U_L(\kappa,\lambda) &=& \frac{\kappa^2}{\tau}\left[H_E+W_2\Delta\right]\,,\\
% %%%%%%%%%%
% U_T(\kappa,\lambda) &=& 2\tau H_M+W_2\Delta\,,\\
%%%%%%%%%%
\widetilde{U}_L(\kappa,\lambda) &=& \frac{\kappa^2}{\tau}
                         \left[\widetilde{H}_E+\widetilde{W}_2\Delta\right]\,,\\
%%%%%%%%%%
\widetilde{U}_T(\kappa,\lambda) &=& 2\tau\widetilde{H}_M+\widetilde{W}_2\Delta\,,\\
%%%%%%%%%%
\widetilde{U}_{T'}(\kappa,\lambda) &=& \widetilde{H}_A(1+\widetilde{\Delta})\,.
\ea
Additionally,
\ba
% H_E &=& Z(G_E^p)^2 + N(G_E^n)^2\,,\\
% %%%%%%
% H_M &=& Z(G_M^p)^2 + N(G_M^n)^2\,,\\
% %%%%%%
% W_2 &=& \frac{1}{1+\tau}\left[H_E+\tau H_M\right]\,,\\
%%%%%%
\widetilde{H}_E &=& ZG_E^p\widetilde{G}_E^p + NG_E^n\widetilde{G}_E^n\,,\\
%%%%%%
\widetilde{H}_M &=& ZG_M^p\widetilde{G}_M^p + NG_M^n\widetilde{G}_M^n\,,\\
%%%%%%
\widetilde{W}_2 &=& \frac{1}{1+\tau}\left[\widetilde{H}_E+\tau\widetilde{H}_M\right]\,,\\
%%%%%%
\widetilde{H}_A &=& 2\sqrt{\tau(\tau+1)}\left[ZG_M^pG_A^{e,p}\right. \nonumber\\
		&+& \left. N G_M^nG_A^{e,n}\right]\,,
\ea
where $Z$ and $N$ represent the proton and neutron number of the target nucleus, respectively.
Finally,
\ba
\Delta &\equiv& \xi_F(1-\psi^2)\left[\frac{\sqrt{\tau(\tau+1)}}{\kappa}\right.\nonumber\\ 
       &+&\left. \frac{1}{3}\xi_F\frac{\tau}{\kappa^2}(1-\psi^2)\right]\,,\\
%%%%%%%
\widetilde{\Delta} &\equiv& \frac{1}{2\kappa}
                           \sqrt{\frac{\tau}{1+\tau}}\xi_F(1-\psi^2)\,,\\
%%%%%%%%
{\cal D} &\equiv& 1 + \frac{1}{2}\xi_F(1+\psi^2)\,.
\ea

% \linespread{0.5}

% \small

\bibliographystyle{apsrev4-1}

\bibliography{bibliography}

\end{document}